\documentclass[preprint2]{aastex}

\newcommand{\um}{$\mu$m}
\newcommand{\brgamma}{Br$\gamma$}
\newcommand{\kms}{km\thinspace s$^{-1}$}

\def\degr{\hbox{$^\circ$}}
\def\arcmin{\hbox{$^\prime$}}
\def\arcsec{\hbox{$^{\prime\prime}$}}
\def\utw{\smash{\rlap{\lower5pt\hbox{$\sim$}}}}
\def\udtw{\smash{\rlap{\lower6pt\hbox{$\approx$}}}}

\def\Lsun{\hbox{\it L$_\odot$}}

\def\Msun{\hbox{\it M$_\odot$}}
\def\Minit{\hbox{\it M$_{\rm initial}$}}

\def\J{\hbox{\it J}}
\def\H{\hbox{\it H}}
\def\K{\hbox{\it K}}

\def\Mk{\hbox{\it M$_{\rm K}$}}
\newcommand{\Ks}{{\it K$_{\rm s}$}}

\newcommand{\Aks}{{\it A$_{\rm K_{\rm s}}$}}
\newcommand{\Ak}{{\it A$_{\rm K}$}}

\def\BCK{\hbox{\it BC$_{\rm K}$}}

\def\simgr{\mathrel{\hbox{\rlap{\hbox{\lower4pt\hbox{$\sim$}}}\hbox{$>$}}}}

\shorttitle{Galactic stellar candidate clusters. Near-infrared spectra.}
\shortauthors{Messineo et al.}

\begin{document}


\title{ Near-infrared spectra of Galactic stellar clusters detected 
on Spitzer/GLIMPSE images. \footnote{Based on observations
collected at the European Southern Observatory 
(ESO Programme 079.D$-$0807).} }


\author{Maria~Messineo\altaffilmark{1}, 
	Ben~Davies\altaffilmark{1},
	Valentin~D.~Ivanov\altaffilmark{2},
        Donald~F.~Figer\altaffilmark{1}, 
	Frederic~Schuller\altaffilmark{4},
 	Harm~J.~Habing\altaffilmark{3},
	Karl~M.~Menten\altaffilmark{4},
	Monika~G.~Petr-Gotzens\altaffilmark{5}
  }

\email{messineo@cis.rit.edu}

\altaffiltext{1}{Chester F. Carlson Center for Imaging Science, Rochester Institute
   of Technology, 54 Lomb Memorial Drive, Rochester, NY 14623-5604, United
   States}
\altaffiltext{2}{European Southern Observatory, Ave. Alonso de Cordova 3107, Casilla 19, Santiago 19001, Chile}
\altaffiltext{3}{Leiden Observatory, P.O. Box 9513, 2300 RA Leiden, the Netherlands}
\altaffiltext{4}{Max-Planck-Institut f\"ur Radioastronomie, Auf dem H\"ugel 69, D-53121 Bonn, Germany}
\altaffiltext{5}{European Southern Observatory, Karl Schwarzschild-Strasse 2, D-85748 Garching bei Munchen, Germany}


%

\begin{abstract}   
We present near-infrared spectroscopic observations of
massive stars in three stellar clusters located in the direction of the inner
Galaxy. One of them, the Quartet, is a new discovery while the other two were
previously reported as candidate clusters identified on mid-infrared Spitzer
images (GLIMPSE20 and GLIMPSE13). Using medium-resolution 
(R$=900-1320$) \H\ and \K\ spectroscopy, we firmly establish the nature of the
brightest stars in these clusters, yielding new identifications of an early WC
and two Ofpe/WN9 stars in the Quartet and an early WC star in GLIMPSE20. We
combine this information with the available photometric measurements from 2MASS,
to estimate  cluster masses, ages, and distances.  The presence of several
massive stars places the Quartet and GLIMPSE20 among the small sample of known
Galactic stellar clusters with masses of a few $10^3$ \Msun, and ages 
from 3 to 8 Myr. We estimate a distance of about 3.5 kpc for
Glimpse 20, and  6.0 kpc for Quartet.
The large number of giant stars identified in GLIMPSE13 
indicates that it is another massive ($\sim 6500$ \Msun) cluster, but older, 
with an age between 30 and 100 Myr, at a distance of about 3 kpc.

\end{abstract}


\keywords{stars: evolution --- infrared: stars }



\section{Introduction}  Stellar clusters are important tools of
Galactic astronomy because they enable measurements of the location and the
motion of spiral arms, of the Galactic rotation curve, and of variations of
metallicity across the Galaxy. Clusters also are excellent laboratories for
testing stellar evolution predictions because they are ensembles of coevally
formed stars that span a range of stellar masses.  Our understanding of stellar
clusters can be used to provide constraints for  models of stellar
evolutionary stages of massive stars (M$_{\sc *}$$\geq 10{\rm~M}_\odot$), e.g.\
Wolf-Rayet (WR) stars, blue supergiants (BSGs), yellow supergiants (YSGs), and
red supergiants (RSGs). Massive stars are of particular interest since they are
not only supernovae progenitors, but are also among the brightest stars in
external galaxies, and contribute to the  chemical enrichment of the
interstellar medium. Massive stars  are typically found only in massive young
stellar clusters (M$_{\sc tot}$ $\gtrsim 10^3{\rm ~M}_\odot$) due to  their
short life span and to the steep  initial mass function.

The current census of Galactic young stellar  clusters is highly incomplete beyond a
few kiloparcsec from the  Sun because of interstellar extinction. So far, clusters
have been detected that reside in  the near side of the Galaxy, with a few
exceptions, e.g. W49  \citep{homeier05}. We probably know only about 5\% of the young
clusters. We expect about 50,000 stellar clusters over the entire Galactic plane when
considering the population of 1700   clusters optically detected, which are mostly
within 3 kpc from the Sun \citep{dias02}, and assuming a uniform distribution over a
disk of 17 kpc. \citet{dias02} provide us with the most complete  compilation of known
clusters detected at optical wavelength. A study of the completeness of cluster
detections, and  cluster parameters, is, however, missing.  Using the cluster density
derived by \citet{lamers05} with the smaller, but homogeneous, list of  500 clusters
in the solar neighbor \citep{kharchenko05} we infer a similar total number of
clusters. 

Only a dozen clusters  with a mass larger than M$_{\sc tot}$$>10^4 M_\odot$ are
known \citep[][]{figer08,borissova08}. From this number, we extrapolate 200
clusters with masses M$_{\sc tot}$$>10^3 M_\odot$ by assuming  a power  law with
an index $\alpha=-2$ as a cluster mass function.   This small fraction of known
clusters may not be representative of  the whole sample. 

Infrared searches for clusters in the Two Micron All Sky Survey (2MASS)
\citep{skrutskie06} and the Galactic Legacy Infrared Mid-Plane Survey
Extraordinaire (GLIMPSE) \citep{benjamin03} have resulted in 1500  additional
new candidate clusters
\citep[e.g.][]{bica03,mercer05,froebrich07,ivanov02,borissova03}.  Many 
candidate clusters could be spurious detections due to the patchiness of
interstellar extinction; some apparent over-density in stellar counts may simply
correspond to low extinction regions.  \citet{froebrich07} point out that only
50\% of their detections are likely to be real. Deep  near-infrared photometry
and spectroscopy are needed to confirm candidate  clusters.

Clearly, we need to improve the census of clusters in the Milky Way,  as well as
confirm new candidate clusters.  Then we will be able to address several
issues concerning the Galactic cluster population, such as the cluster
initial mass function, the dynamical evolution of young clusters  into globular
clusters, and the paucity of stellar clusters within 3 kpc from the Galactic
Center.

In this paper, we analyze near-infrared spectroscopic observations of two
stellar clusters, number \#20 and \#13  from the list of candidate clusters
detected by  \citet{mercer05} in GLIMPSE images, and of a new cluster, which we
also detected in GLIMPSE images. Hereafter we refer to cluster \#20 of Mercer et al.
as GLIMPSE20, and to cluster \#13 as GLIMPSE13. We name the new cluster ``the
Quartet'', because it is characterized by a tight grouping of 4 
mid-infrared bright stars. This  work presents the massive stellar content of these
three clusters based on spectroscopic findings and existing 2MASS data.  A more
detailed study of each individual cluster will be presented in future papers.   
We describe  observations and data reduction in Section 2, and spectral
classification in Section 3. We present  an analysis and discussion of the 
Quartet cluster in Section 4, of GLIMPSE20 in Section 5, and of GLIMPSE13 in
Section 6.  Section 7 contains a summary  and discussion.

\begin{deluxetable}{lrrr}
\tablewidth{0pt}
\tablecaption{\label{obs.list} List of observed stellar clusters.}
\tablehead{
\colhead{Cluster}& 
\colhead{RA(J2000)}& 
\colhead{DEC(J2000)} & 
\colhead{radius(\arcsec)} 
}
\startdata
Quartet   & 18 36 17.0 &  $-$07 05 02 &   45  \\
GLIMPSE20 & 19 12 23.7 &  $+$09 57 10 &   60  \\ 
GLIMPSE13 & 18 53 53.0 &  $+$00 37 41 &   45  \\
\enddata
\tablecomments{
Coordinates indicate the cluster center, which are calculated as a 
flux weighted centroid of a 2MASS \Ks-band smoothed image. 
The apparent cluster radius is also reported.} 
\end{deluxetable}

\section{Observations and data reduction} We selected candidate clusters from the list
of \citet{mercer05} by searching  GLIMPSE images for small groupings of bright stars
with no associated nebulosity.  In addition, we included a new cluster, the Quartet,
that we discovered while inspecting GLIMPSE images.  The observed targets are listed in
Table \ref{obs.list}.  We obtained near-infrared spectra of the brightest  stars in
the clusters (Table \ref{table.targets})   with the SofI spectrograph at the ESO NTT 
on July 4, 2007. We first used the instrument  in imaging mode to acquire the field and 
position the target into the slit.  Then, we  adjusted the slit orientation to maximize
the number of observed stars. Typically, we extracted three or four spectra per slit. 
The integration time ranged from 10 to 180 s in \Ks-band and from 10 to 240 s in
\H-band. Spectra, taken with a long-slit (1\arcsec$\times 290$\arcsec), have a
resolving power R$\sim$1320 in the \Ks-band (from 2.0 to 2.31 $\mu$m) and R$\sim$900
in the \H-band (from 1.4 to 1.8 $\mu$m). We observed two stars with \Ks$<5$ mag 
(GLIMPSE13 \#1 and GLIMPSE20 \#1) with a narrow slit (0.6\arcsec$\times 290$\arcsec),
boosting the resolution up by a factor of 1.67. For each star and filter, we obtained
four spectra  in an ABBA sequence by nodding the telescope along the slit between
exposures. During the second half of the night, clouds  hampered our observations, so
only \Ks-band spectra were taken.

Data was reduced with the Image Reduction and Analysis Facility  (IRAF\footnote{IRAF is
distributed by the National Optical Astronomy Observatories, which are operated by the
Association of Universities for Research in Astronomy, Inc., under cooperative agreement
with the National Science Foundation.}) software. We removed  the sky  by subtracting
pairs of subsequent images, and  flat-fielded the images using a combined set of
dark-subtracted dome flats. We extracted the individual spectra using an aperture of 10
pixels ($2\rlap{.}$\arcsec88), and subtracted sky residuals  using the average value of
regions selected on both sides of the aperture. The spectra were wavelength calibrated
using arc observations. Finally, we combined  the four spectra of each source.

Stars of spectral types G1V and G2V were observed  close to the airmass of the program
stars as telluric standards; their spectra were taken, and reduced in the same manner as
those of the target stars.  We divided the reference spectra by a solar spectrum to obtain
a combined instrumental and atmospheric response function. In the \H-band, the use of
G-type stars as telluric standards resulted in a non-optimal correction (residuals are
stronger between 1.55 and 1.65 \um). This is because spectra of G-type stars have
intrinsic features at $> 5$ \% of the continuum level.

\section{Spectral classification of member stars}   To perform a spectral
classification, we compared the spectra of the detected stars  with \K-band and
\H-band spectral atlases
\citep[e.g.][]{hanson96,hanson98,hanson05,meyer98,ivanov04, alvarez00, blum96,
kleinmann86,wallace96}.

{\bf Late-type stars}: Spectra of late-type stars contain several diagnostic features in
the \K-band  \citep[e.g.\ a CO overtone band, Mg I, Ca I, and Na I lines,][]{ivanov04}, as well as
in the \H-band  \citep[e.g. lines from Mg I, Al I, OH, and  second-overtone CO band heads
are detected,][]{meyer98}. K and M type stars are easy to identify because of their deep CO
band head absorption at 2.29 \um.  The CO absorption strength increases with: decreasing
effective temperature, and increasing luminosity \citep{mcwilliam84}. The H$_2$O absorption
(at 1.35--1.5, 1.7--2.1, 2.7 \um) increases with decreasing temperature,  and  decreasing
luminosity.  Water absorption is a distinctive feature of pulsating variable giants,  
while  almost absent in static giants and in RSGs. We estimated effective temperatures  of
late type stars following  the method of \citet{davies07}, i.e.\  by comparing the
measurements of equivalent widths of the CO band head at 2.29 \um\ with those of template
stars \citep{kleinmann86,wallace96}.

{\bf Early-type stars:} Early-type stars can be identified by the detection of hydrogen
(H) and helium (He) lines.  Several metal lines are also detected at infrared
wavelengths:  for example, the CIV triplet at 2.069, 2.078, and 2.083 $\mu$m; and the
broad emission feature found at 2.116 $\mu$m, identified as N III or CIII emission
\citep{hanson96,hanson98,hanson05,kleinmann86,meyer98}.  WR stars are also rich in
nitrogen and carbon lines \citep{figer97,crowther06,morris97}. 

We used the ratio between the emission line at 2.11\um\  (CIII+HeI) and the CIV
line at 2.08 \um\   to classify carbon rich WR stars.

{\small

\begin{deluxetable}{llllrrrll}
\tablewidth{0pt}
\tablecaption{\label{table.targets} List of observed stars.}
\tablehead{
\colhead{Cluster}& 
\colhead{ID}& 
\colhead{RA}& 
\colhead{DEC} & 
\colhead{\J} & 
\colhead{\H} & 
\colhead{\Ks} &  
\colhead{Spec.Type} &  
\colhead{Member} 
}
\startdata
Quartet&1 & 18:36:17.29 & $-07$:05:07.3 & 10.34 $\pm$  0.02 &  8.40 $\pm$  0.02 &  7.38 $\pm$  0.02    &   Ofpe/WN9 & yes \\  
       &2 & 18:36:16.69 & $-07$:04:59.5 & 10.57 $\pm$  0.02 &  8.66 $\pm$  0.03 &  7.58 $\pm$  0.02    &   Ofpe/WN9 & yes\\  
       &3 & 18:36:18.70 & $-07$:05:05.7 & 13.17 $\pm$  0.04 &  9.74 $\pm$  0.06 &  7.93~$\pm $\nodata  &     K--M  & no \\  
       &4 & 18:36:19.21 & $-07$:04:02.3 & 11.18 $\pm$  0.03 &  8.99 $\pm$  0.02 &  8.02 $\pm$  0.01    &   K--M  & no \\  
       &5 & 18:36:16.33 & $-07$:05:17.0 & 13.22 $\pm$  0.03 & 10.14 $\pm$  0.02 &  8.09 $\pm$  0.03    &   WC9 & yes \\  
       &6 & 18:36:17.55 & $-07$:04:58.8 & 11.90 $\pm$  0.03 & 10.06 $\pm$  0.03 &  9.18 $\pm$  0.04    &   OB & yes   \\  
       &7 & 18:36:16.88 & $-07$:05:14.0 & 12.27 $\pm$  0.03 & 10.21 $\pm$  0.03 &  9.19 $\pm$  0.03    &   OB & yes  \\
       &8 & 18:36:18.08 & $-07$:04:42.4 & 13.34 $\pm$  0.03 & 10.86 $\pm$  0.02 &  9.75 $\pm$  0.03    &   K--M  & no \\ 
       &9 & 18:36:21.03 & $-07$:03:10.2 & 14.35 $\pm$  0.04 & 11.47 $\pm$  0.02 & 10.12 $\pm$  0.02    &   OB & yes   \\  
\tableline
GLIMPSE20 &   1 & 19:12:23.72 & $+09$:57:08.1 &  7.46 $\pm$  0.02 &  5.81 $\pm$  0.03 &  4.89 $\pm$  0.02&GI & yes   \\  
	  &   2 & 19:12:24.77 & $+09$:56:38.3 & 10.07 $\pm$  0.02 &  7.71 $\pm$  0.04 &  6.55 $\pm$  0.02&K--M & no  \\  
	  &   3 & 19:12:23.59 & $+09$:57:38.0 & 11.47 $\pm$  0.03 &  9.89 $\pm$  0.04 &  8.97 $\pm$  0.03& OB  & yes \\  
	  &   4 & 19:12:23.70 & $+09$:57:19.4 & 11.70 $\pm$  0.08 & 10.32 $\pm$  0.08 &  9.10 $\pm$  0.05&K--M & no  \\  
	  &   5 & 19:12:25.09 & $+09$:57:47.0 & 12.65 $\pm$  0.03 & 10.25 $\pm$  0.04 &  9.16 $\pm$  0.04&K--M & no  \\  
	  &   6 & 19:12:24.14 & $+09$:57:29.1 & 12.04 $\pm$\nodata& 10.38 $\pm$  0.04 &  9.19 $\pm$  0.04&WC   & yes \\  
	  &   7 & 19:12:21.11 & $+09$:56:45.0 & 14.13 $\pm$  0.02 & 10.83 $\pm$  0.02 &  9.24 $\pm$  0.03&K--M & no  \\  
	  &   8 & 19:12:23.14 & $+09$:56:46.3 & 11.51 $\pm$  0.02 & 10.20 $\pm$  0.02 &  9.56 $\pm$  0.02&OB &  yes  \\  
	  &   9 & 19:12:21.46 & $+09$:56:53.2 & 11.99 $\pm$  0.03 & 10.67 $\pm$  0.03 &  9.96 $\pm$  0.03&OB &  yes  \\  
	  &  10 & 19:12:18.66 & $+09$:55:53.4 & 14.53 $\pm$  0.03 & 11.46 $\pm$  0.03 & 10.03 $\pm$  0.02&K--M & no  \\  
	  &  11 & 19:12:23.37 & $+09$:57:33.7 & 13.34 $\pm$  0.06 & 11.48 $\pm$  0.07 & 10.57 $\pm$  0.07&K--M & no  \\  
\tableline	
 GLIMPSE13 &   1 & 18:53:52.49 & $+00$:39:31.3 &  7.32 $\pm$  0.02 &  4.29 $\pm$  0.21 &  2.67 $\pm$  0.25&   M0I/M7III\\ 
	   &   2 & 18:53:52.40 & $+00$:40:17.2 & 11.51 $\pm$  0.02 &  7.85 $\pm$  0.03 &  5.89 $\pm$  0.02&   K5III\\	
	   &   3 & 18:53:53.89 & $+00$:37:12.2 & 10.27 $\pm$  0.03 &  8.10 $\pm$  0.03 &  7.13 $\pm$  0.02&   M1III\\	
	   &   4 & 18:53:53.18 & $+00$:38:00.1 & 10.59 $\pm$  0.03 &  8.56 $\pm$  0.03 &  7.61 $\pm$  0.02&   K5III\\	
	   &   5 & 18:53:52.05 & $+00$:37:57.7 & 11.35 $\pm$  0.03 &  8.96 $\pm$  0.03 &  7.79 $\pm$  0.02&   M0III\\	
	   &   6 & 18:53:53.59 & $+00$:37:08.4 & 11.08 $\pm$  0.03 &  9.03 $\pm$  0.03 &  7.90 $\pm$  0.05&   K4III\\	
	   &   7 & 18:53:53.71 & $+00$:38:05.5 & 10.89 $\pm$  0.03 &  8.86 $\pm$  0.03 &  7.95 $\pm$  0.02&   K5III\\	
	   &   8 & 18:53:53.14 & $+00$:37:22.9 & 11.22 $\pm$  0.03 &  8.95 $\pm$  0.03 &  7.95 $\pm$  0.02&   K5III\\	
	   &   9 & 18:53:53.59 & $+00$:37:46.4 & 11.04 $\pm$  0.04 &  8.95 $\pm$  0.03 &  7.98 $\pm$  0.02&   K5III\\	
	   &  10 & 18:53:53.06 & $+00$:37:05.6 & 11.10 $\pm$  0.03 &  9.09 $\pm$  0.03 &  8.19 $\pm$  0.02&   K5III\\	
	   &  11 & 18:53:54.47 & $+00$:37:23.0 & 11.72 $\pm$  0.03 &  9.41 $\pm$  0.03 &  8.38 $\pm$  0.02&   K5III\\	
	   &  12 & 18:53:45.60 & $+00$:36:25.7 & 11.49 $\pm$  0.02 &  9.44 $\pm$  0.03 &  8.46 $\pm$  0.03&   K1III\\	
	   &  13 & 18:53:52.44 & $+00$:37:21.7 & 12.07 $\pm$  0.03 & 10.11 $\pm$  0.03 &  9.08 $\pm$  0.03&   G9III\\	
	   &  14 & 18:53:53.22 & $+00$:38:27.6 & 12.39 $\pm$  0.03 & 10.49 $\pm$  0.03 &  9.70 $\pm$  0.03&   G9III\\	
	   &  15 & 18:53:55.94 & $+00$:37:20.3 & 13.82 $\pm$  0.07 & 11.27 $\pm$  0.04 & 10.08 $\pm$  0.03&   K2III\\	
\enddata
\tablecomments{For each star, number designations and coordinates
(J2000) are followed by  \J, \H, and \Ks\ measurements  from 2MASS,
spectral classification from this work, and comments.}
\end{deluxetable}
}
\section{Quartet cluster}   
\begin{figure}[!]
\begin{center}
\resizebox{0.95\hsize}{!}{\includegraphics[angle=0]{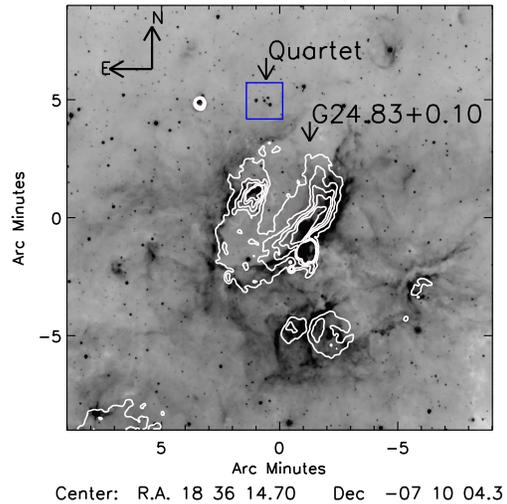}}
\caption{\label{bubblequartet} GLIMPSE 8 \um\ image of the region
around the Quartet cluster (square). The image is displayed in a
linear scale, the maximum is set to 250 MJy sr$^{-1}$.  The image is
centered on the adjacent HII region G024.83+00.10. Contours are
from the 20~cm map \citep{helfand06}. Contour levels, in mJy
beam$^{-1}$, are: 3, 5, 7, 9, 11. The beam size is 24\arcsec $\times$
18\arcsec, FWHM, and the position angle of the major axis is along the
North-South direction. }
\end{center}
\end{figure}

The newly  discovered 
stellar cluster is located on the Galactic plane at (l,b)=(24\fdg90,+0\fdg12).
A quartet of bright stars, with \Ks\ between 7 and 8 mag, dominate the
mid-infrared cluster light (Fig.\ \ref{bubblequartet}).   The cluster is in a
crowded region, is quite concentrated and small,  with $\sim$45\arcsec\
diameter as measured on \Ks\ images, and  lies at the border of a GLIMPSE image.
Likely this is why previous searches  missed it \citep[e.g.][]{mercer05}.

The Quartet is projected on the border of a mid-infrared bubble (see Fig.\
\ref{bubblequartet}). This bubble  (N36) was detected by \citet{churchwell06},  and
coincides with the HII region G024.83$+$00.10.   Radio recombination line observations 
of this HII region \citep{kantharia07} yielded an LSR velocity,  V$_{\rm LSR}$, of
$106\pm2$ \kms;  a near kinematic heliocentric distance of $6.1\pm0.6$ kpc
\citep{brand93}; and a  far distance of $9.2\pm0.1$ kpc.

\begin{figure}[!]
\begin{center}
\resizebox{0.95\hsize}{!}{\includegraphics[angle=0]{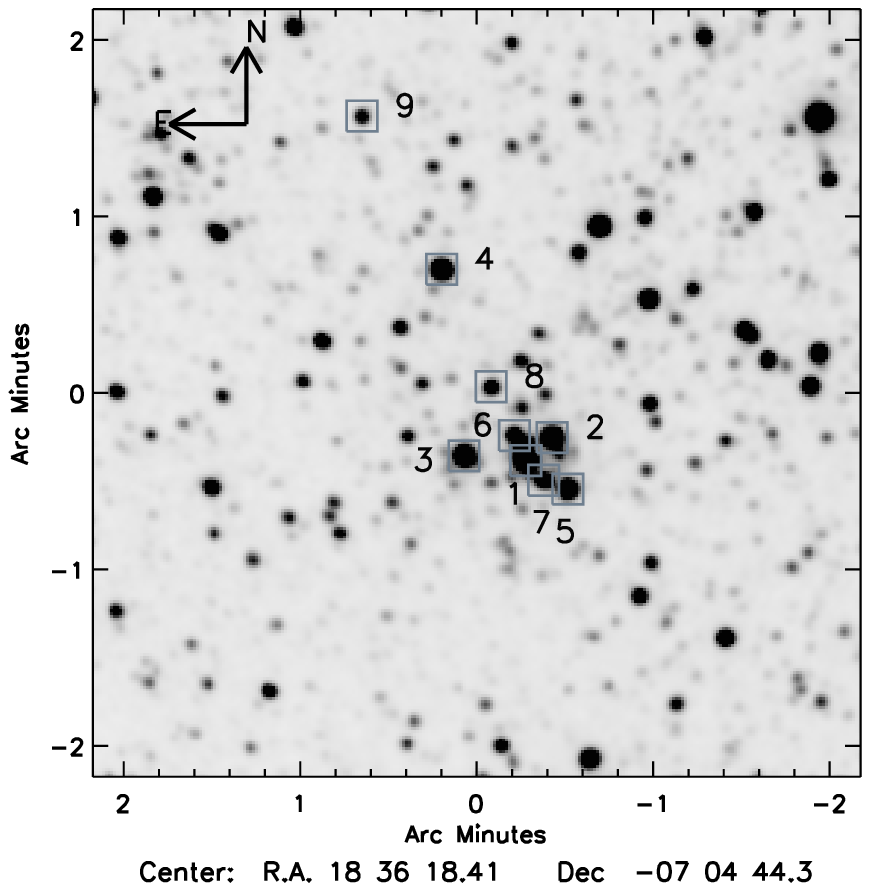}}
\caption{\label{quartet} 2MASS \Ks-band image of the Quartet cluster.
Squares indicate the stars with spectra. Number designations are from
Table \ref{table.targets}. }
\end{center}
\end{figure} 

\subsection{Spectra} 
\begin{figure}[!]
\begin{center}
\resizebox{0.95\hsize}{!}{\includegraphics[angle=0]{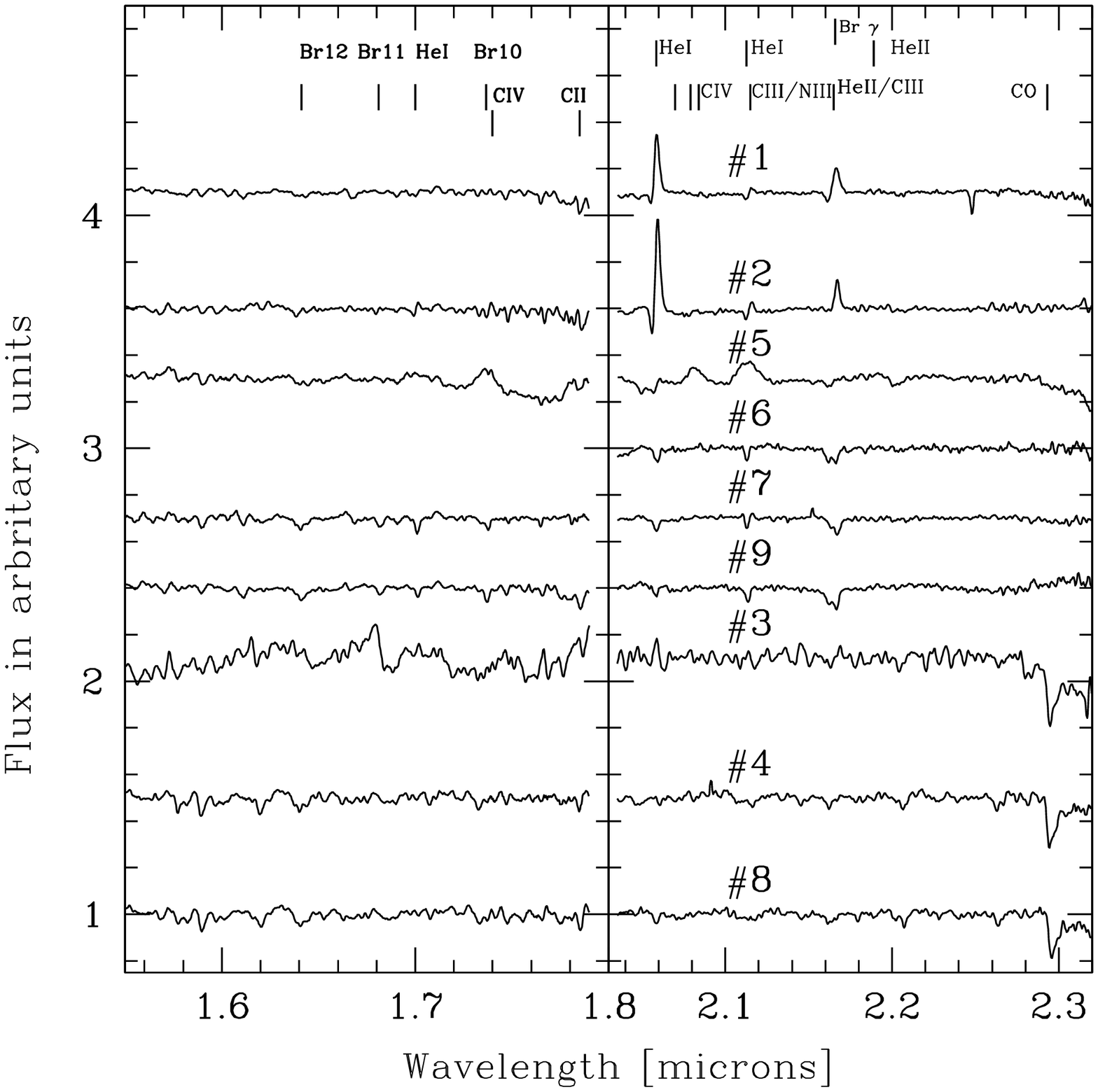}}
\end{center}
\begin{center}
\resizebox{0.7\hsize}{!}{\includegraphics[angle=0]{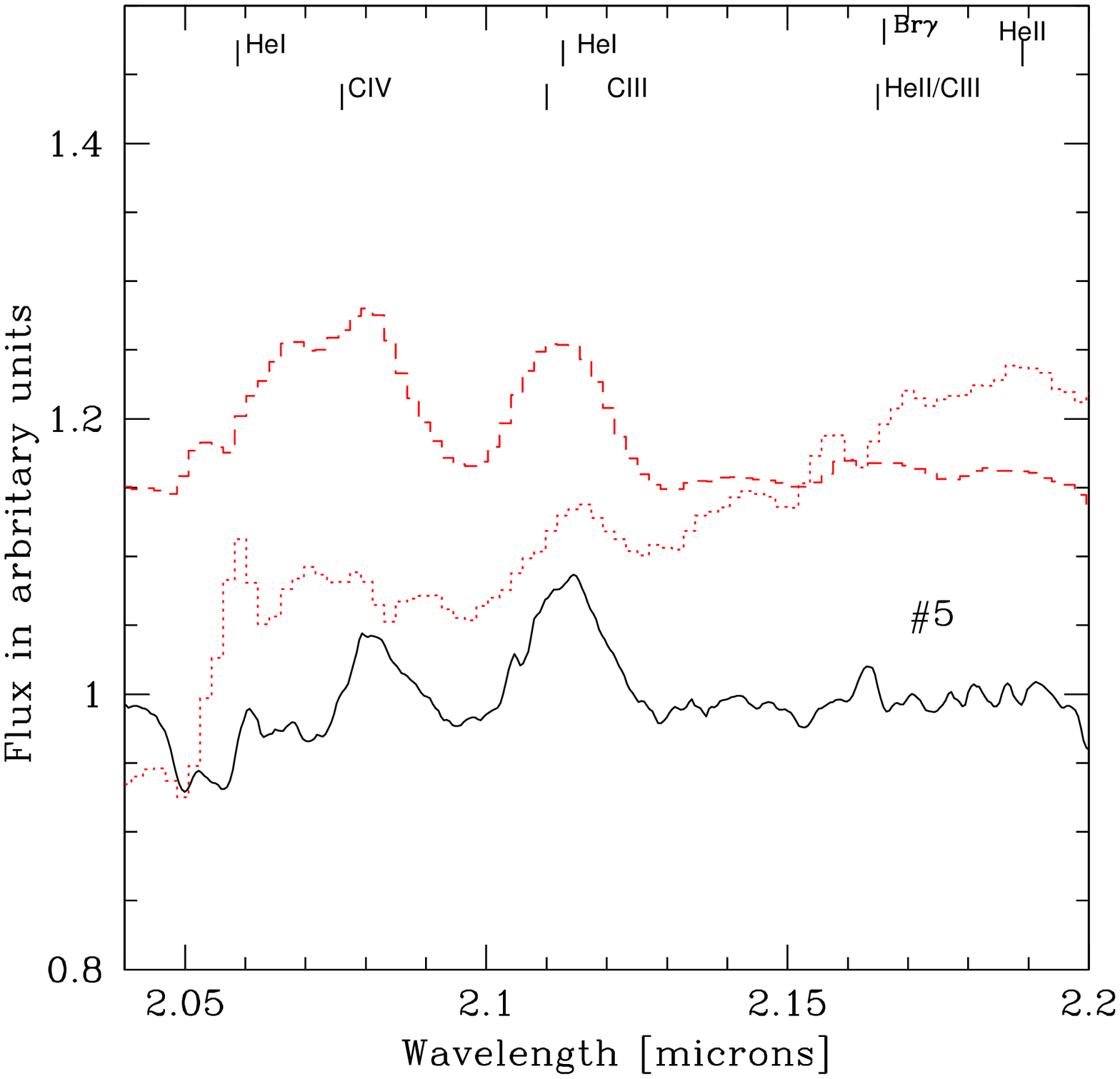}}
\end{center}
\caption{\label{quartet.spectra} {\bf Top:}\H\ and \K-band spectra of stars
observed towards the Quartet cluster. {\bf Bottom:} \K-band spectrum of
the new WC9 star (continuum line). As a comparison, spectra of a WC8 
(dashed line) and a WC9  (dotted line) stars from the atlas of \citep{figer97}
 are also shown.}
\end{figure}

Three slit positions were used to observe the brightest
stars of the cluster.  A total of 9 spectra with a signal-to-noise of the
continuum above 40  were extracted (Table \ref{table.targets}  and Fig.\
\ref{quartet.spectra}). 

\begin{itemize} \item The spectra of stars \#1 and \#2 show strong  HeI lines at
2.058 $\mu$m and \brgamma\ lines in emission, and a weak line at 2.11 $\mu$m
(HeI/NIII/CIII). The P-Cygni profile of the lines indicates the presence of 
expanding envelopes.  The FWHM values of the HeI line at 2.058 $\mu$m are 411 and 430
\kms, respectively. The detected lines and measured FWHM values agree with those of
known Ofpe/WN9 stars \citep{blum95,figer99, martins07}.  Ofpe/WN9 stars have been
also detected in the Quintuplet \citep[see Fig. 3,][]{figer95}, and in the Galactic
center \citep[see Figs. 2 and 4,]{martins07}. The HeI (2.058)/\brgamma\ ratios we obtained are 1.4 and
1.6, i.e. higher than 0.9 as found in the Quintuplet stars, but smaller than 2, the
typical ratio found in Galactic center HeI emission line stars
\citep{blum95,figer95}. Star \#1 and \#2 are likely to be Ofpe/WN9 stars. They are
located within 7\arcsec\ of the adopted cluster center.

\item Star \#5 shows several broad emission features. From a comparison with 
the atlases of WR stars by \citet{crowther06},  \citet{figer97}, and
\citet{morris97}, we identified HeI/CII at 1.700 \um, CIV at 1.74 \um, CII at
1.785 \um, CIV at 2.08 \um, CIII at 2.115 \um.  We concluded that star  \#5 is a
carbon WR star.  The ratio between the equivalent width of the CIV line at 2.078
\um\ ($11.9\pm1.2$) and that of CIII at 2.116 \um\ ($17.2\pm0.8$) is smaller
than unity, indicating a spectral subtype WC9 \citep[e.g.][]{crowther06}. The
star is 17\arcsec\ away from the cluster center.

\item Stars \#6, \#7, and \#9 show HeI lines at 2.058 \um,   HeI lines at 2.11
\um, and \brgamma\ lines in absorption in the \K-band. The \H-band spectra of
stars \#7 and \#9 show HeI lines at 1.70 \um, as well as Br10, Br11 and Br12
lines, in absorption.  These spectra are similar to those of stars in the
O9--B1 class of \citet{hanson96}.  Stars \#6 and \#7 are within 11\arcsec\ from the
cluster center, while star \#9 is 127\arcsec\ away.

\item Star \#3 ( \Ks$=7.9$ mag) is a late-type star as indicated by the presence of a strong CO
band head at 2.29\um\ in absorption. Since a RSG star is expected to be brighter in
\Ks-band than an early-type supergiant of the same cluster (Fig.\
\ref{quartet.cm}, \Ks$=9.2$ mag),  we concluded that star \#3 is a giant star,
unrelated to the Quartet.  Stars \#4 and \#8 (\Ks=8.02 mag and \Ks=9.75 mag),
which also show CO band heads in absorption, lie at larger distances from the
cluster center. They were not our primary targets, but just happened to
fall on the slit. \end{itemize}

\subsection{Color-magnitude diagram}  
\begin{figure}[!t]
\begin{center}
\resizebox{0.95\hsize}{!}{\includegraphics[angle=0]{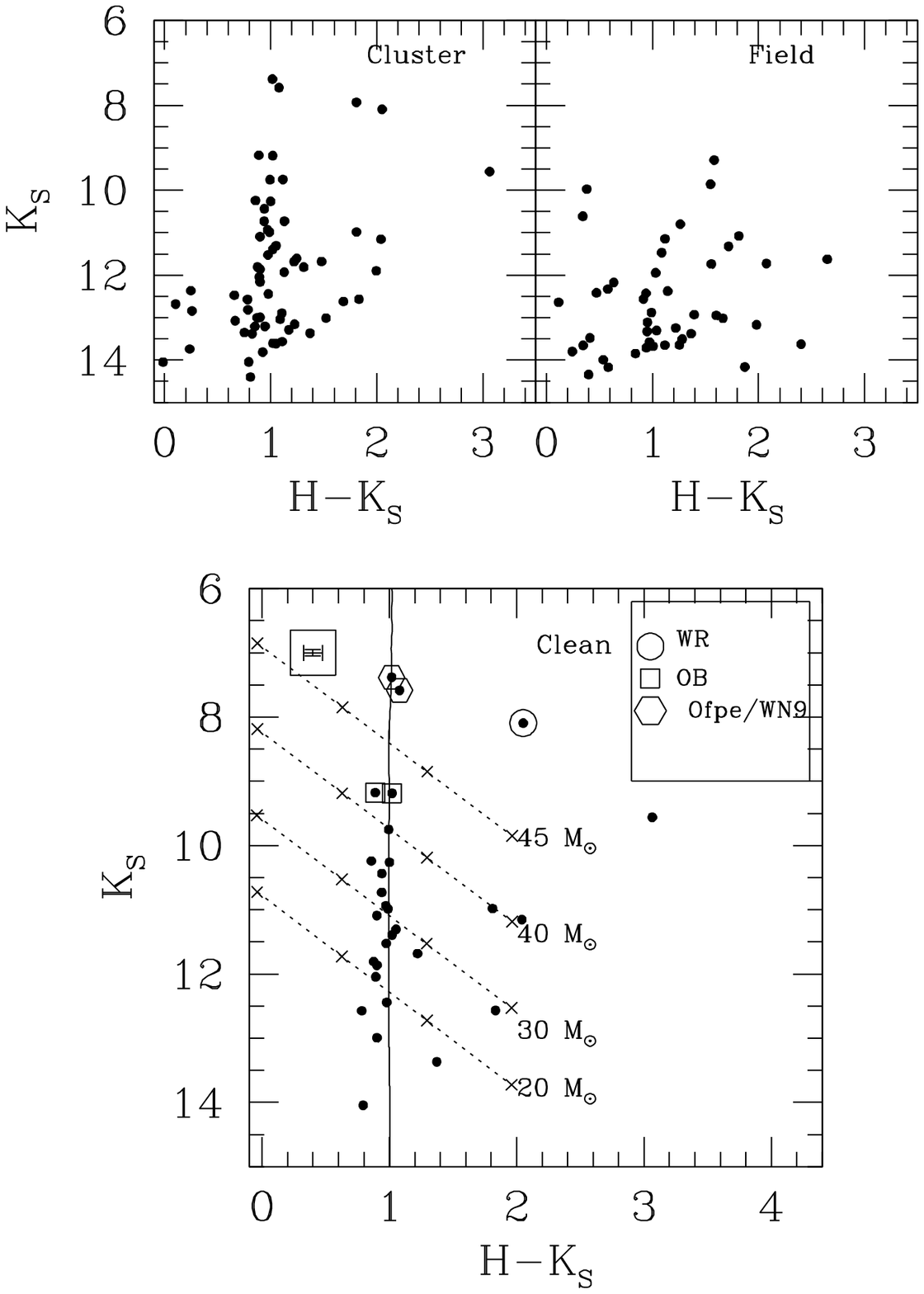}}
\end{center}
\caption{\label{quartet.cm} 2MASS (\H$-$\Ks) versus \Ks\ diagrams of the 
Quartet cluster (upper-left panel) and of a comparison field (upper-right
panels) with identical area (radius is 45\arcsec). The decontaminated diagram is
shown in the lower panel. The solid  line indicates an isochrone with solar
abundance and an age of 4 Myr \citep{lejeune01}, which has been shifted to a
distance of 6.1 kpc and reddened to  \Aks$=1.6$ mag using the extinction law
by \citet{messineo05}.  The locations of stars on the isochrone with initial
masses of 20, 30, 40 and 45 \Msun\ and an interstellar extinction \Aks\ from 0
to 3 mag are indicated by dashed lines.  Squares indicate the location of the
spectroscopically confirmed OB stars: an open circle shows the WR star, 
and hexagons the Ofpe/WN9 stars. The typical error for the plotted 2MASS 
sources is plotted within the top left box.}
\end{figure}

The (\H$-$\Ks) versus \Ks\ diagram of
2MASS point sources within 45\arcsec\ of the cluster center is shown in Fig.\
\ref{quartet.cm}.  A well defined cluster sequence appears at $H-$\Ks $=\sim 1$
mag and $12 > $ \Ks$ > 7$ mag.  To identify the cluster sequence and to estimate
the percentage of contamination from background and foreground stars (field
stars), we performed a statistical decontamination. We considered as a field
region an annulus of equal area with an inner radius of 2\arcmin; we calculated
the numbers of stars per bins of ($H-$\Ks)  color and \Ks\ magnitude in both
cluster and field regions;  then, we randomly subtracted from each cluster bin a
number of stars equal to that of the corresponding field bin.  
Incompleteness differences between the field and cluster  only
affect objects at the fainter end of the CMD, i.e. at \Ks$>~12.5$ mag.
In our subsequent analysis we are mainly concerned with objects brighter
than \Ks$>~12.5$ mag, for which we consider the same level of completeness
for cluster and field respectively. The resulting number of 2MASS
candidate cluster members (\Ks$<12.5$ mag) is 27.

\begin{figure}[!]
\begin{center}
\resizebox{0.7\hsize}{!}{\includegraphics[angle=0]{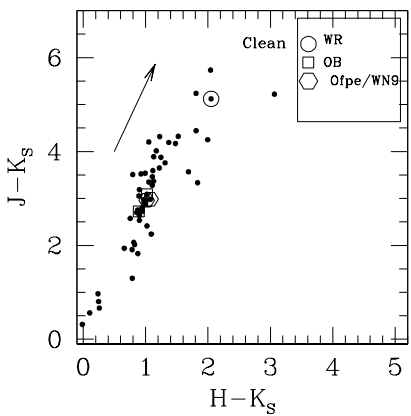}}
\end{center}
\caption{\label{quartet.cc} 2MASS (\H$-$\Ks) versus (\J$-$\Ks) diagrams of the 
Quartet cluster. Symbols are as in Fig.\  \ref{quartet.cm}.}
\end{figure}

\subsection{Discussion}

\subsubsection{Massive stars and  cluster sequence}   \label{quartetdiscussion}  
The  Quartet cluster hosts several massive stars. We spectroscopically detected:
one WC9 star, two Ofpe/WN9, and three OB stars (two within 45\arcsec). 
These stars are among the
brightest cluster members in the \Ks-band. In  Figure \ref{quartet.cm} we present
a 2MASS color-magnitude diagram  (CMD); we over-plot an isochrone of 4 Myr, 
solar  abundance, without rotation and with high mass-loss, from the Geneva group
\citep{lejeune01}, shifted to the spectro-photometric distance of 6.3 kpc, as
determined below. An interstellar extinction value of \Aks$= 1.6\pm0.4$ mag was
obtained by fitting the color of the observed cluster sequence with the
theoretical isochrone. This measurement is insensitive to age since the isochrones
are almost vertical lines.



On a 2MASS $H-Ks$ versus $J-Ks$  diagram the cluster appears  as a concentration
of stars with  $H-K_{\rm s}=\sim 1$ mag. The colors of the WR star are redder
than that measured toward the cluster sequence (see Fig.\ \ref{quartet.cc}), 
suggesting the presence of circumstellar dust.  By assuming absence of
circumstellar dust, we estimated a reddening of $E(J-K_{\rm s})=4.89$ mag and
$E(H-K_{\rm s})=1.79$ mag, and an extinction of \Aks$=2.65\pm0.04$ mag, using
intrinsic $J-Ks=0.23$ mag, $H-Ks$=0.26 mag \citep{crowther06}, and the infrared
extinction law by \citet{messineo05}. This confirms that  the reddening of the
WR star is higher than that measured toward the cluster sequence (\Aks$=1.6$
mag), and confirms the presence of circumstellar dust. The formation of dust
around WR stars can  be explained by means of colliding winds, which, in turn,
require a binary system; therefore, the most likely scenario  is that dusty WC9
stars are binary stars \citep{crowther06}. Only about forty other WC9 stars are
known in the Galaxy at present \citep[][]{hopewell05}.

The detection of two Ofpe/WN9 stars makes the Quartet a particularly interesting cluster. 
These stars are rare objects with initial masses between 25 and 60 \Msun\ in transition
from the red supergiant stage to the WR stage (WN8) \citep{martins07, smith04}. The
Quintuplet and the Galactic center clusters are the only other Galactic stellar clusters
that are known to contain Ofpe/WN9 stars \citep{figer99,martins07}.

Absolute magnitudes of Ofpe/WN9  and  WC9 stars are quite uncertain. Therefore,
we derived a spectro-photometric distance from the OB  stars, \#6 and \#7.  
We assumed the intrinsic colors and bolometric corrections of  blue supergiant
stars  given by \citet{bibby08}, with spectral types from O9.5 to B3, and the
extinction law by \citet{messineo05}. A spectro-photometric distance of
$6.3\pm2.0$ kpc is obtained, where  the error is due to uncertainty of the 
spectral types \citep{messineo07a}.  This range of distances is compatible with
the hypothesis that the cluster is close to the HII region G$024.83+00.10$,
if the near-side kinematic distance of $6.1 \pm 0.6$ kpc is assumed
\citep{brand93}.

The presence of evolved stars, such as the WR star and the two Ofpe/WN9, suggests that the
cluster is older than $\sim$3 Myr. Since WR stars are present in populations younger than
$\sim$8 Myr  \citep[\Minit$ > 22$ \Msun,][]{meynet05},  the cluster must also be younger
than $\sim$8 Myr. The lack of RSG stars also constrains the age to $< \sim 8$ Myr
\citep{figer06}. This  yields an age range of 3-8 Myr.

For the two Ofpe/WN9 stars, we derived luminosities of $\sim 22\pm15 \times
10^5$ \Lsun\ and  $\sim 19\pm12 \times 10^5$ \Lsun\ by assuming a distance of
$6.3\pm2.0$ kpc, an average interstellar extinction \Aks$=1.6\pm0.4$ mag, and a
bolometric correction \BCK$=-2.9\pm1.0$ mag \citep{martins07,najarro97}.  
These luminosities are similar to those quoted by \citet{martins07} for the
Ofpe/WN9 in the Galactic center (1--6 $\times 10^5$ \Lsun), as well as to those
of Ofpe/WN9 stars detected in the Quintuplet cluster \citep[28--45$ \times 10^5$
\Lsun,][]{figer99}.

We counted at most $22\pm5$ stars with masses above 20 \Msun\ (\Ks $\leq12.13$
mag) on the decontaminated color-magnitude diagram, assuming a distance of
$6.3\pm2.0$ kpc, an extinction \Aks$=1.6\pm0.4$ mag, and the 4 Myr   theoretical
isochrone, without rotation, with high-mass loss rates, and with solar abundance 
\citep{lejeune01}. If we use an isochrone of 7.9 Myr, the number of stars above
20 \Msun\ (\Ks $=10.89$ mag) is at least $13\pm8$.  By assuming a Salpeter
initial mass function \citep{salpeter55}, and extrapolating  it to  0.8 \Msun, we
concluded that the cluster has a total  mass between 1300 and 5200 \Msun.  The
uncertainty is due to uncertain age and distance.

\subsubsection{Luminosity and Ionizing Flux} The Quartet cluster appears in
projection adjacent to the HII region G024.83$+$00.10.  We investigated the
possibility that the stellar cluster could be the heating and ionizing source
of the HII region.

Two IRAS sources, IRAS 18335$-$0713A and IRAS 18335$-$0711,  coincide with the position of
this HII region.  We found that the cloud has a far infrared luminosity of L$_{IR } = 2.8
\times 10^5$ \Lsun\  by using the IRAS flux densities, the kinematic distance of 6.1 kpc,
and the relation between the IRAS flux densities and the integrated far infrared
luminosity \citep{lumsden02}. \citet{kantharia07} found that the radio continuum emission
of the HII region at 49 and 20 cm  is consistent with optically thin free-free emission,
and estimated that a number of Lyman continuum photons N$_{Ly}$$=2.3 \times 10^{49}$
s$^{-1}$ is required to maintain the ionization.


The cluster luminosity's lower limit was estimated by  assuming a distance of
6.1 kpc as for the HII region, together with an average extinction of \Aks$=1.6$
mag, and \K-band bolometric corrections from Table 3 of \citet{figer99}. We
added the luminosity of the brightest/massive stars  ($> 20$ \Msun);  we assumed
that stars without known spectral types were  OB stars. Following
\citet{figer99}, a conservative average bolometric correction of \BCK$=-2.0$ mag
was assumed to account for spectral types later than O9. We obtained a cluster
luminosity of $7.3 \times 10^6$ \Lsun. Furthermore, we estimated  a Lyman
continuum flux N$_{Ly}$$=5 \times 10^{49}$ s$^{-1}$ by integrating the
contributions from the brightest stars.  Using an isochrone with solar
abundance and an  age of 4 Myr, we estimate that the TO is located at (\Ks=11.5
mag) and  Teff=34000 K.  This temperature corresponds to that of an O8.5 stars
\citep{martins05}.  The two spectroscopically detected OB stars (\Ks$=9.2$ mag)
are  similar to stars in the O9-B1 class of \citep{hanson96}. The majority of
the brightest stars are  expected to be evolved stars with spectral type from O8
to B1. We therefore  assumed an average O9.5 type for the 15 stars  with \Ks$ <
9.2$ and \Ks $> 11.5$ mag, and adopted Lyman fluxes for a specific spectral type from
\citet{martins07} and \citet{martins05}.

The number of Lyman continuum photons needed to ionize the HII regions appears
comparable to that produced by the stellar cluster,  when assuming that the
cluster is at the same distance as the cloud and an average spectral type of O9.5. 
The cluster could plausibly contribute to the ionization of the HII region.

\section{GLIMPSE20} 

\begin{figure}[!]
\begin{centering}
\resizebox{0.95\hsize}{!}{\includegraphics[angle=0]{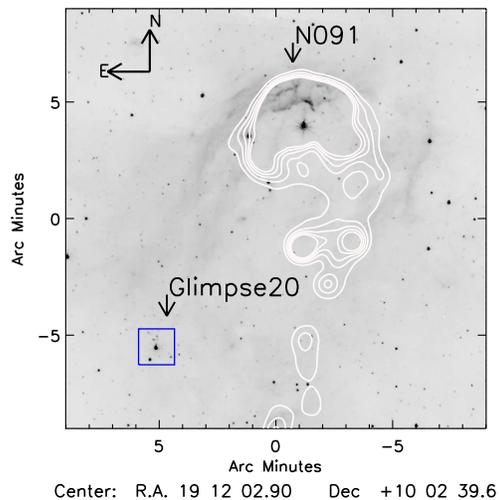} }
\caption{\label{bubbleglimpse20} GLIMPSE 8 \um\ image of the
GLIMPSE20 cluster (square) and of the adjacent mid-infrared
bubble N091.  The image is displayed in a linear scale, and the
maximum is set to 250 MJy sr$^{-1}$.  Overlaid contours are from the
NVSS 21~cm data \citep{condon98}.  Contour levels, in mJy beam$^{-1}$,
are: 3, 5, 7, 9, 11.  The beam size is 45\arcsec\ FWHM.}
\end{centering}
\end{figure} 

The GLIMPSE20 cluster was discovered by \citet{mercer05}. It appears as a
concentration of bright stars (with a radius of about 1\arcmin) at both
mid-infrared (GLIMPSE) and near-infrared (2MASS) wavelengths. The cluster is
located at (l,b)=(44\fdg16,$-$0\fdg07) on the border of the mid-infrared bubble
N091 (Fig.\ \ref{bubbleglimpse20}), for which \citet{churchwell06} list a
kinematic distance of 4.9 kpc (near) or 7.3 kpc (far). The morphology of the
mid-infrared bubble and the radio continuum emission, i.e. the lack of radio
continuum emission in the cluster's vicinity, suggests that the cluster is 
not the ionizing source.

\subsection{Spectra} 

\begin{figure}[!] 
\begin{center}
\resizebox{0.95\hsize}{!}{\includegraphics[angle=0]{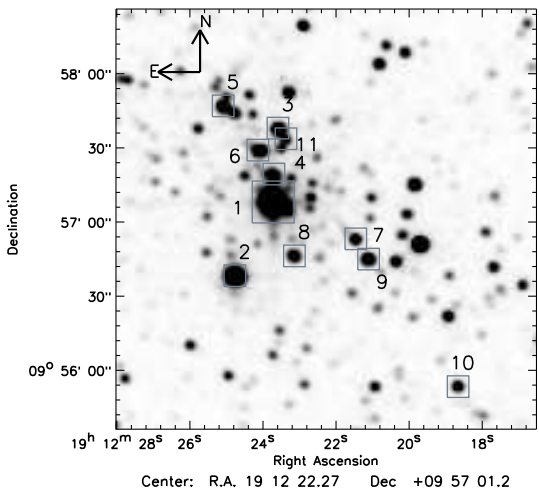} }
\end{center} 
\caption{\label{gl20map} 2MASS \Ks-band image of the GLIMPSE20
cluster.  Squares indicate the stars with spectra. 
Number designations are from Table \ref{table.targets}.}
\end{figure}

\begin{figure}[!]
\begin{center}
\resizebox{0.95\hsize}{!}{\includegraphics[angle=0]{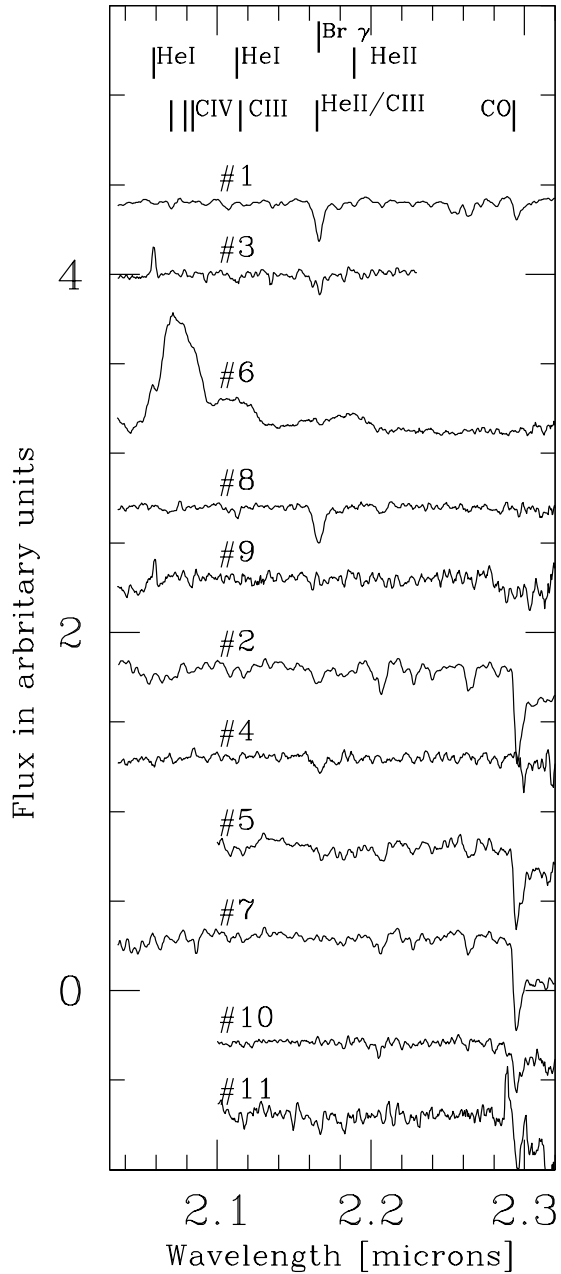}}
\end{center}
\caption{\label{gl20spectra} { \H\ and \K-band spectra of the stars 
detected towards GLIMPSE20. Regions with unreliable telluric subtraction,
and uncertain continuum level are omitted.  } }
\end{figure}

\begin{figure}[!] \begin{centering}
\resizebox{0.95\hsize}{!}{\includegraphics[angle=0]{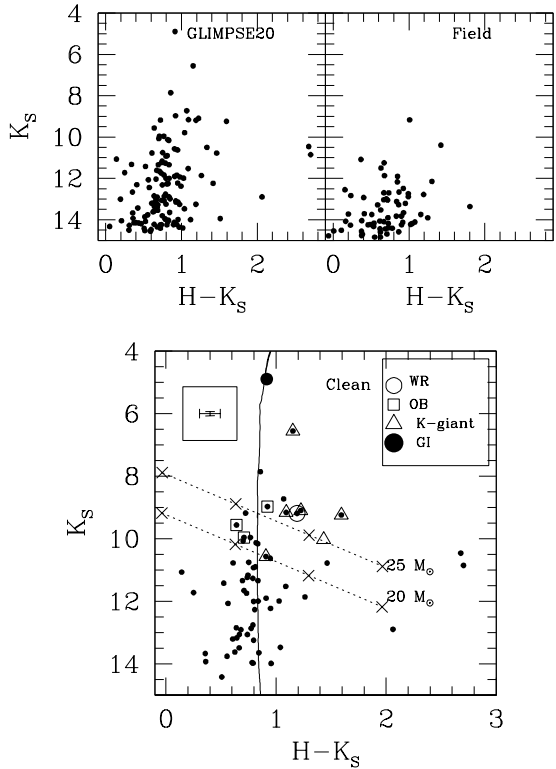}} 
\end{centering} 
\caption{\label{fig:cmdgl20.ps}  2MASS (\H$-$\Ks) versus \Ks\ diagrams of the GLIMPSE20
cluster (upper-left panel) (radius is 59\arcsec) and of a comparison annular field
(upper-right panels) with identical area and an inner radius of 2\arcmin. The
decontaminated diagram is shown in the lower panel. The solid  line indicates an isochrone
with solar abundance and an age of 6.3 Myr \citep{lejeune01}, which has been shifted to a
distance of 3.8 kpc and reddened to an \Aks$=1.3$ mag using the extinction law by
\citep{messineo05}.  The locations of stars on the isochrone with initial masses of 20 and
25 \Msun\ and an interstellar extinction \Aks\ from 0 to 3 mag are indicated by dashed
lines.  Squares indicate the location of the spectroscopically confirmed OB stars, an open
circle shows the WR star,  a filled circle the yellow supergiant, and triangles the late
type stars. Symbols without corresponding central dots are stars located outside the
central area of 59\arcsec\ radii, or stars which were statistically removed by the
decontamination algorithm. The typical errors for the plotted 2MASS 
sources is plotted within the top left box.}  
\end{figure}

Four different slit positions were used to observe the brightest stars in the 
GLIMPSE20 cluster. A total of 11 spectra were extracted with signal-to-noise
above forty. The targeted stars are shown in  Fig.\ \ref{gl20map}, while the
spectra are shown in Fig.\ \ref{gl20spectra}.

\begin{itemize} \item The bright star \#1 (\Ks$=4.8$ mag) is located in the
central region of the cluster. The detection of a weak CO band head at 2.29
$\mu$m and of \brgamma\ line in absorption indicates a late F or early G type.

\item  Star \#6 shows strong broad lines in emission from CIV, CIII and HeII,
which indicate a carbon WR star.  We have simultaneously fitted two Gaussians 
to the blended lines from CIV (peak is at 2.073 \um)  and CIII (peak is at
2.114 \um), and measured an equivalent width of $776\pm59$ $\AA$ and $261\pm39$
$\AA$, respectively.  A comparison with the \K-band WR star atlas of
\citet{figer97} indicates that star \#6 has a spectral-type between WC4 and
WC7.  Large intrinsic variations exist in equivalent widths  within each
spectral subclass, which hamper a more precise classification based only on
\K-band features.

\item Star \#8 shows a \brgamma\ line and a weak HeI (2.11 \um) line,  both in
absorption.  A comparison with the atlas by \citet{hanson96} indicates an O9--B2
star.  Star \#3 also shows a \brgamma\ line and a weak HeI  at 2.11 \um\ line in
absorption; the HeI line at 2.058 $\mu$m seems to be in emission, which suggests an
O9--B2 supergiant. The spectrum of star \#9 is quite noisy; however, the lack of
strong CO absorption suggests  a spectral type earlier than K.

\item Stars \#2, \#4, \#5, \#7, \#10, and \#11 show CO in absorption,
suggesting low effective temperature. This is in agreement with the locations
of these stars on the 2MASS CMDs. Therefore, they are field stars unrelated to
the cluster. \end{itemize}

\subsection{Color-magnitude diagram and discussion}  2MASS CMDs of the 
GLIMPSE20 cluster are shown in Fig.\ \ref{fig:cmdgl20.ps}.  A cluster sequence
appears at (\H$-$\Ks) $ \sim0.8$ mag.

The  WR star, star \#6, is located within a bright group of stars at \Ks$\sim9$
mag.  For star \#6, we assumed absence of circumstellar  envelope, and
used intrinsic (\J$-$\Ks), (\H$-$\Ks), and \Mk\
values for early-type WC stars as given by \citet{crowther06}. Using these
values and the extinction law by \citet{messineo05}, we derived an interstellar
extinction of \Aks $=1.0\pm0.2$ mag and a distance of $3.8\pm1.3$ kpc.
The derived extinction is consistent with that towards the OB stars.

Star \#1, a star (\Ks$=4.8$ mag) four magnitudes brighter than other candidate
cluster members, dominates the infrared cluster light.  The combination of
2MASS magnitudes and spectral type (G-type) indicates that \#1 is a yellow
supergiant (YSG).  A foreground G0--2 V star and a G0--2 III star can be ruled out
because their intrinsic magnitudes \citep[e.g.][]{wainscoat92} would require a
distance of only 11 to 40 pc, but a large interstellar extinction.  We derived
\Aks$=1.2$ mag \citep{messineo05}, which is similar to the value obtained for
the WC star, by comparing the observed colors ((\J$-$\Ks) $=2.57$ mag and
(\H$-$\Ks) $=0.91$ mag) with intrinsic colors from \citet{koorneef93}.
Since the luminosities of known yellow supergiants range from $1.0 \times 10^5$
to $6.3 \times 10^5$ \Lsun\ \citep{smith04}, the photometric properties of the
newly identified YSG star yield a spectrophotometric distance between 3.8 and
9.0 kpc. When assuming the distance derived for the WR star, we obtained an
absolute \Ks$=-9.2$ mag and a luminosity of $1.0 \times 10^5 $ \Lsun\
\citep[\BCK$=1.4$ mag,][]{koorneef93}. For the YSG in the RSGC1 cluster,
\citet{davies08} report an absolute \Ks$=-10.07$ mag and a luminosity of $2.3
\times 10^5 $ \Lsun;   this luminosity agrees well with that of the YSG in
GLIMPSE20 within the uncertainty of  the correction for interstellar
extinction. 


For the BSG star \#3 (\Ks = 8.9 mag), we used intrinsic infrared magnitudes
from \citet{bibby08}, and estimated an extinction \Ak$ = 1.5$ mag and a
distance between 3.9 and 6.6 kpc.  Star \#8 is a fainter star (\Ks = 9.6 mag); 
its early-type and  interstellar extinction (\Aks$= 1.1$ mag) suggest a
relation to the cluster.  We derived a distance of only 2.0 kpc, when adopting
the intrinsic magnitudes for a  dwarf O9.5V star \citep{martins06}; of 3.9 kpc,
assuming a giant O9.5III star; of 6.3 kpc, assuming a O9.5I star. Star \#8 is
likely to be a giant.

Spectroscopically detected late-type stars are located on the right side
of the cluster sequence.  Stars \#4, \#5, \#7, \#10, and \#11 are fainter than the
detected OB stars, suggesting  they are background stars that are unrelated to
the cluster.  Star \#2 (\Ks$=6.5$ mag) is also most likely a non-member since
it is fainter than the YSG star.

Within errors, the interstellar extinction and spectrophotometric distances of
the YSG, the WR star, and the early-type stars are consistent; his
confirms the reality of the stellar cluster, which has
a most likely distance of  3.8 and 5.1 kpc.

Evolutionary models, without rotation and with high mass-loss rates, and
solar metallicity from \citet{lejeune01}, predict the presence of YSG/RSG
stars in a population with an age greater than 6--7 Myr, while WR stars
have ages younger than 8 Myr (\Minit$ = 22$ \Msun, \Ks$<10.26$).  We derived
a cluster age between 6 and 8 Myr by assuming coevality between the yellow
supergiant and the WR star.

After performing a statistical field decontamination of the CMD, we found
$17\pm6$ candidate cluster stars more massive than 20 \Msun, where the error
is due to the uncertainties in distance and age.   The total cluster mass was
estimated in the same way as for the Quartet  cluster (see Sec.
\ref{quartetdiscussion}), and we  obtain  $3400\pm1300$ \Msun.

\section{GLIMPSE13}

\begin{figure}[!]
\begin{centering}
\resizebox{0.95\hsize}{!}{\includegraphics[angle=0]{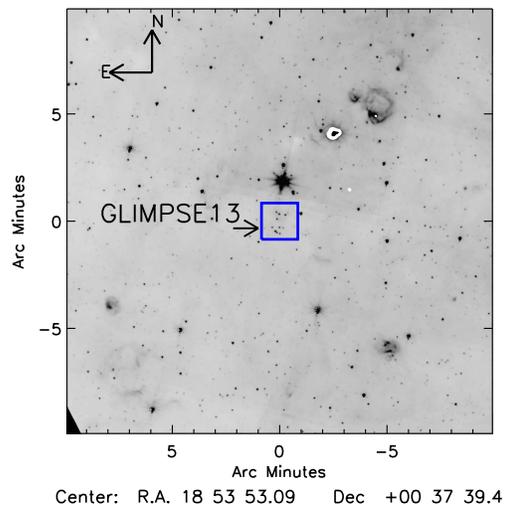}}
\caption{\label{bubbleglimpse13} GLIMPSE 8 \um\ image of the region
around the GLIMPSE13 cluster (square). The image is displayed in a
linear scale, the maximum is set to 250 MJy sr$^{-1}$.  White contours are
from the 20~cm map \citep{helfand06}. Contour levels, in mJy
beam$^{-1}$, are: 3, 5, 7, 9, 11. The beam size is 24\arcsec $\times$
18\arcsec, FWHM, and the position angle of the major axis is along the
North-South direction. }
\end{centering}
\end{figure}

\begin{figure}[!]
\begin{centering}
\resizebox{0.95\hsize}{!}{\includegraphics[angle=0]{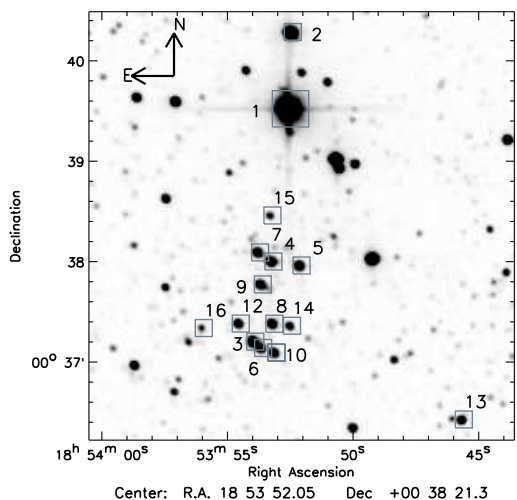}}
\end{centering}
\caption{\label{gl13map} 2MASS \Ks-band image of the GLIMPSE13
cluster.  Squares indicate the observed stars. Number designations 
are from Table \ref{table.targets}.  }
\end{figure}

\begin{figure}[!]
\begin{center}
\resizebox{0.95\hsize}{!}{\includegraphics[angle=0]{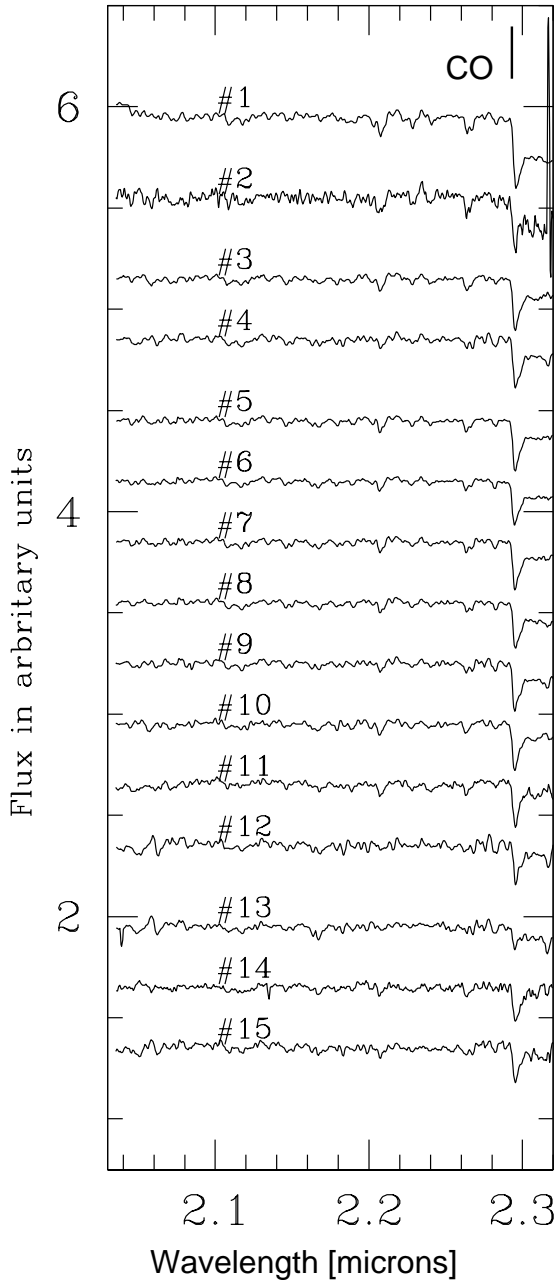}}
\end{center}
\caption{\label{gl13spectra} \K-band spectra
of stars detected towards GLIMPSE13.}
\end{figure}

\begin{figure}[!]
\begin{centering}
\resizebox{0.95\hsize}{!}{\includegraphics[angle=0]{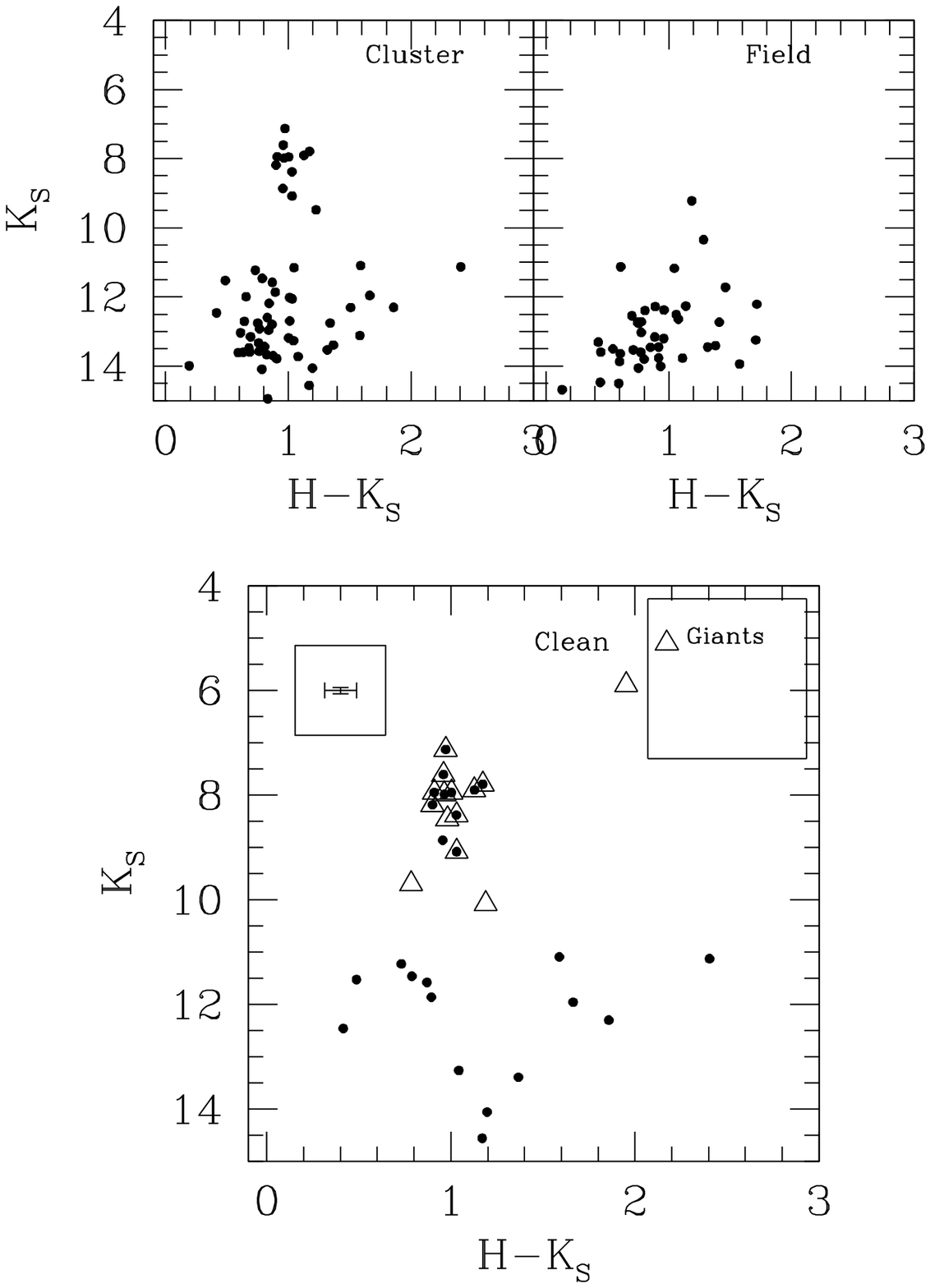}}
\end{centering}
\caption{\label{fig:cmdgl13.ps} 2MASS (\H$-$\Ks) versus \Ks\ diagrams
of datapoints within 45\arcsec\ of the center of GLIMPSE13
(upper-left  panel) and of a comparison field (upper-right panel)
with identical size. Triangles indicate the location of
spectroscopically detected late type stars. Several of them are located
outside the area of 45\arcsec radius. The typical errors for the plotted 2MASS 
sources is plotted within the top left box.}
\end{figure}

The  GLIMPSE13 cluster is located on the Galactic plane at 
(l,b)=(33\fdg79,$-$0\fdg24) (see Figs. \ref{bubbleglimpse13} and
\ref{gl13map}).  A 2MASS (\H$-$\Ks) versus \Ks\ diagram is shown in Fig.\
\ref{fig:cmdgl13.ps}; the diagram shows a distinct clump of bright stars with
similar colors, which were the targets of our spectroscopic observations. A
chart  of GLIMPSE13 stars, for which near-infrared spectra were taken,  is
shown in Fig.\ \ref{gl13map}.

\subsection{Spectra} CO band heads were detected in all \K-band spectra (Fig.\
\ref{gl13spectra}).  As shown by \citet{davies07}, the absorption strength of
the CO band head can be used to estimate effective temperatures of cool stars. 
As giants and supergiants follow a different equivalent width EW(CO) versus
temperature relation, the EW(CO) is also a discriminant of luminosity class. We
measured an equivalent width  of the CO band head feature  for each star
between 2.285 \um\ and 2.315 \um, with an adjacent continuum measurement made
at 2.28--2.29 \um. Then, we compared these measurements to those of template
stars \citep{wallace96,kleinmann86}. With the exception of star \#1, we found
equivalent widths (12 -- 30 \AA) typical of giants with a spectral type of
early-M. We found corresponding spectral types from early-K to late-G, when
assuming the stars are supergiants. As yellow supergiants are rarer than red
supergiants, and as the median spectral-type of RSGs in the Galaxy is M2
\citep{elias85,massey03b}, it is more likely the stars in GLIMPSE13 are giants.
Star \#1 has a CO equivalent width too large to be a giant, suggestive of a
higher luminosity class.

\subsection{Color-magnitude diagram and discussion}  Spectroscopic measurements
indicate that the clump of bright stars in GLIMPSE13 is comprised of giants.  A
clump of bright infrared stars appears to be a common feature of simple stellar
populations with ages between 10 Myr and a few hundred Myr.  This clump consists
of RSGs \citep{figer06,davies07} at ages below $<30$ Myr. As the population
increases in age, the initial masses of the clump stars decrease, as do their
luminosity classes.  There is a gap in magnitude between the clump   and the
main sequence, which decreases with time, and remains constant for giant stars
older than about  2 Gyr \citep{figer04}.

\begin{figure}[!] 
\begin{centering}
\resizebox{0.95\hsize}{!}{\includegraphics[angle=0]{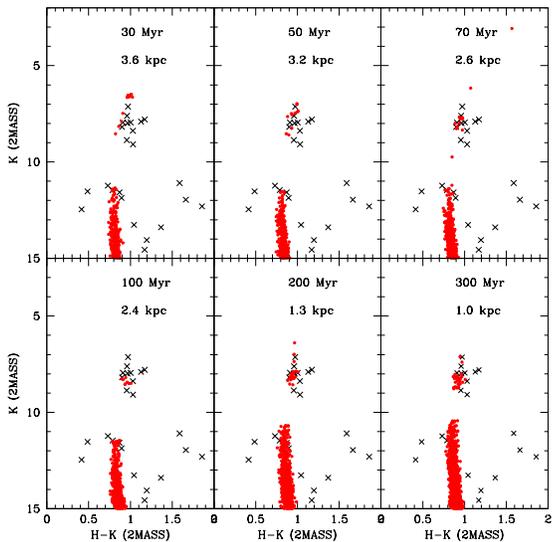}} 
\end{centering}
\caption{\label{models}  2MASS(\H$-$\Ks) versus \Ks\ diagrams of point  sources
within 45\arcsec\ from the cluster center (crosses) are shown. Simulations
(dots) of a simple stellar population of a given age (see label) are also
displayed. Simulated sequences  have been reddened (\Aks $=1.3$ mag) and shifted
in distance in order to match the observed color and magnitude  of clump stars
(see label).  Labels report ages and distances of the simulated stars.} 
\end{figure}

To analyze the observed diagrams of GLIMPSE13, we simulated CMDs for a grid of simple
stellar populations (coeval stars with the same chemical abundance) of different ages. We
used a set of stellar tracks,   with solar composition and with extra core mixing via
overshooting, and a Salpeter initial mass function \citep{cordier07}.  The models are shown
in  Fig.\ \ref{models}, superimposed on the observed CMD. We reddened the model to
\Aks$=1.3$ mag to match the observed 2MASS (\H$-$\Ks), and adjusted the distance of the
simulated population by matching the \Ks-band magnitudes of the clump stars with those of
the observed clump.

In Table \ref{glimpse13.table}  we list the ages and upper masses of each
simulation, as well as the distance obtained by vertically shifting the simulated
diagrams.    The equivalent widths of the CO band heads suggest giants, therefore
the stars must be older than 30 Myr. The models show that for ages of 200--300
Myr, or older, the number of  main sequence  stars ($12<$\Ks$<10$ mag) increases
by a factor three with respect to the clump stars. For an age of 200--300 Myr,
despite the incompleteness of the 2MASS data,  a main sequence   should be
detectable. We have performed simulation with artificial stars and found that
while the field is complete down to \Ks=12.5 mag, the cluster center (radius 60 arcsec)
is  70-85\% complete in this range of magnitude. Therefore, Poisson noise
and incompleteness account for $6\pm6$ cluster stars in this range. 
This suggests that the  GLIMPSE13 cluster has an age between 30 and
100 Myr, with a distance of about $3\pm1$ kpc (see Table \ref{glimpse13.table}). 
The luminous giant stars in GLIMPSE13  have initial masses of $7\pm2$ \Msun\ with
uncertainty due to errors in age/distance. Follow up studies with deeper and
higher-resolution photometric  and spectroscopic  data are needed to further
investigate this cluster and to verify individual memberships.  The number of
observed luminous giant stars corresponds to a total cluster mass of about
$6700\pm2200$ \Msun. Stellar clusters with similar masses are rare in the Galaxy,
because they are located at the top end of the initial cluster mass function of
Galactic open stellar clusters \citep{kroupa02,lamers06}. GLIMPSE13 is the oldest
known  massive cluster.  Therefore, clusters with masses around $10^4$\Msun\  (the
current cluster mass cut-off) were also forming 200-300 Myr ago.

GLIMPSE13 appears as a particularly interesting cluster because it can provide
constraints for stars with 5--10 \Msun, i.e.\ in the crucial mass range for
studying the transition  between massive stars that  evolve into RSGs and
explode as a supernovae, and stars that  evolve into luminous asymptotic giant
branch stars and then as white dwarf. \citep{eldridge04}.  Star \#1 (\Ks$=2.67$
mag) is located within 2\arcmin\ from the cluster center.  The models suggest
that star \#1 could be a massive AGB  star related to the cluster. Follow-up
observations are needed in order to  measure radial velocities, which would
enable us to verify its association  with the cluster.


\begin{table*}[!t]
\caption{ \label{glimpse13.table} Simulations.
Simulations of a simple stellar
population with the BASTI code \citep{cordier07}.
Stars are simulated with mass between 0.8 and 120 M$_\odot$.
The age of the simple stellar population is followed by the number of
clump stars (\Ks$<$10 mag), main sequence stars (10 mag $>$\Ks$>$12 mag), 
heliocentric distance, maximum stellar mass, and cluster mass.}
\begin{center}
\begin{tabular}{rlllll}
\hline
Age   & N$_{\rm star}$ & N$_{\rm star}$   &  Distance & M$_*$ & M$_{\sc tot}$\\
(Myr) & (\Ks$<$10 mag)& 10 $>$\Ks$>$12 &  kpc & \Msun  & \Msun\\
\hline
 30   &12	& 18$\pm$ 7  &	3.6& 8.7 & 6338 $\pm$1873\\
 50   &12       & 12$\pm$ 4  &	3.2& 6.8 & 4576 $\pm$1305\\
 70   &12       & 16$\pm$ 5  &	2.6& 5.9 & 3552 $\pm$ 961\\
100   &12       & 12$\pm$ 4  &	2.4& 5.0 & 2671 $\pm$ 697\\
200   &12       & 32$\pm$12  &	1.3& 4.0 & 1305 $\pm$ 372\\
300   &12       & 41$\pm$14  &	1.0& 3.4 &  ~~933 $\pm$ 257\\
\hline
Observed &12    & 6  &      &\\
\hline
\end{tabular}
\end{center}
\end{table*}

\section{Summary} 

\begin{figure*}[!]  
\begin{centering}
\resizebox{0.95\hsize}{!}{\includegraphics[angle=0]{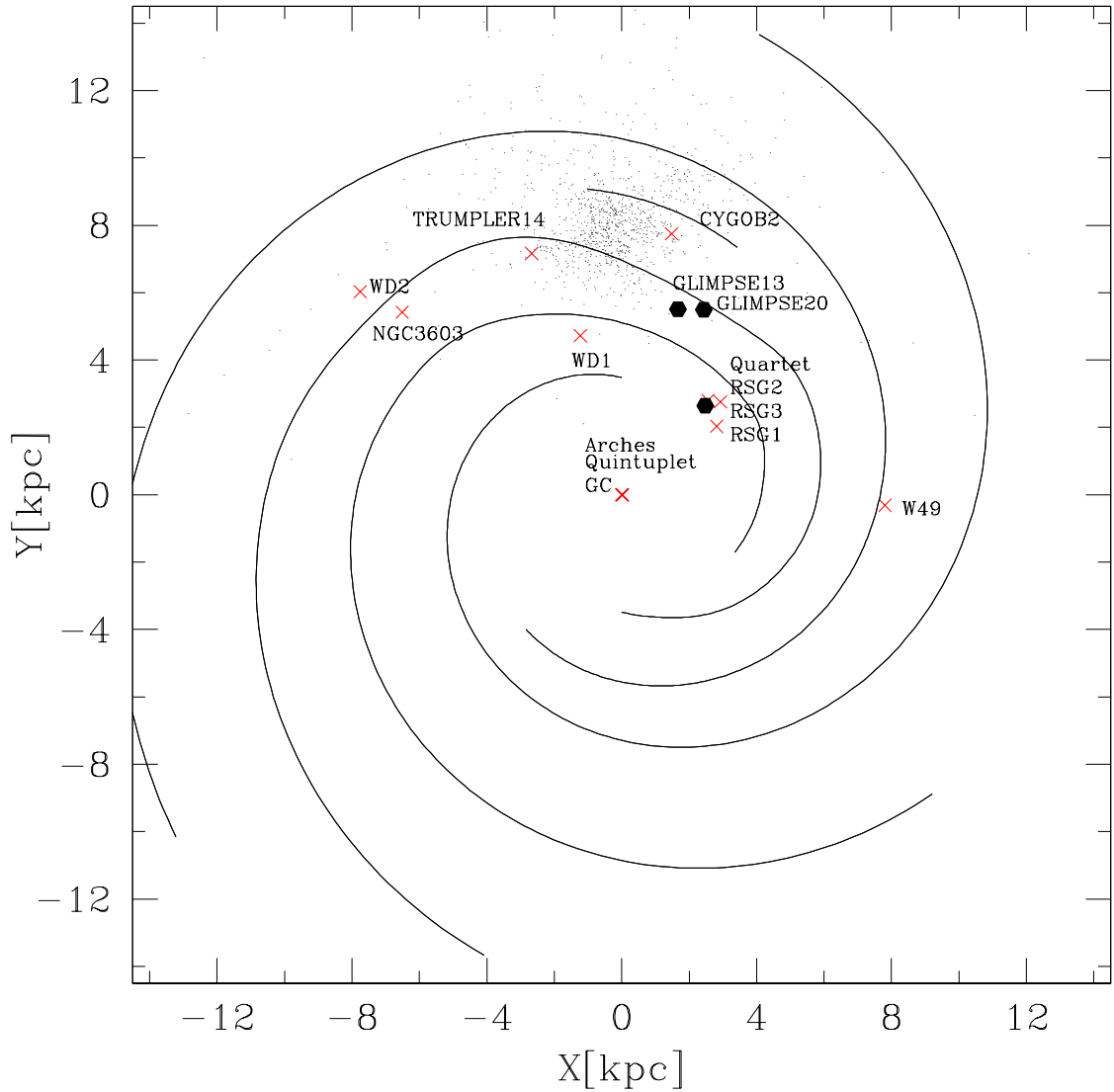}}  
\end{centering}
\caption{\label{xy.ps} Galactic distribution of clusters optically detected (dots)
taken from \citep{dias02}, known massive clusters   (see Table
\ref{table.massive}) are shown with  crosses. The Galactic center is at (0,0)
and  the Sun is at (0,8). The three clusters presented in this work are marked
with hexagons. Spiral arms are from \citet{lazio02}.} 
\end{figure*}

\begin{deluxetable}{lrrrrrl}
\tablewidth{0pt}
\tablecaption{\label{table.massive} Galactic massive clusters ($>10^4$ Msun).}
\tablehead{
\colhead{Cluster}& 
\colhead{Lon [deg]}& 
\colhead{Lat [deg]}& 
\colhead{Distance [kpc]} &  
\colhead{Age [Myr]}& 
\colhead{Mass [$10^3$ \Msun]} & 
\colhead{References}  
}
\startdata
 RSGC2       & 26.2 & 0.0    & $5.8^{+1.9}_{-0.8}$   & $17\pm3$ &$40 \pm10 $     &\citet{davies07}\\
 Westerlund1 &339.5 & $-0.4$ & $3.6\pm0.2$   & $3.6\pm0.7$  &$36 \pm22 $     &\citet{brandner08}\\
 RSGC1       & 25.3 & $-0.2$ & $6.6\pm0.9$   & $12.0\pm2.0$ &$30 \pm10 $     &\citet{davies08}\\
 RSGC3       & 29.2 & $-0.2$ & $6\pm1$       & $18.0\pm2.0$ &$30 \pm10 $     &\citet{clark09}\\
  Arches      &  0.1 &    0.0 & $7.62\pm0.32$\tablenotemark{a}     & $ 2.5\pm0.5$ &$\sim20 $             &\citet{figer08,figer99b}\\
 Quintuplet  &  0.2 &    $-0.1$ & $7.62\pm0.32$\tablenotemark{a}     & $4\pm1$      &  $\sim20$           &\citet{figer08,figer99b}\\
 GC central  &  0.0 &    0.0 & $7.62\pm0.32$\tablenotemark{a}& $6.0\pm2.0$   & $\sim 20$\tablenotemark{b} &\citet{martins07,figer08}\\
             &      &        &               &              &$1,000 \pm 500 $&\citet{schoedel09}\\        
 NGC3603     &291.6 &  $-0.5$& $6.0\pm0.8$   & $<2.5$       &$13\pm3$       &\citet{harayama08}\\
 Trumpler14  &287.4 &  $-0.6$& $\sim 2.8$    & $3.25\pm2.75$&$10\pm1$       &\citet{ascenso07b}\\
 Cyg OB2     & 80.2 &  0.8   & $\sim 1.5$    & $\sim2.5$    &               $\sim 10$\tablenotemark{c}     &\citet{negueruela08}\\ 
 W49A        & 43.2 & 0.0    & $11.4\pm1.2$  & $1.2\pm1.2$  &$\sim 10$&\citet{homeier05}\\  
 Westerlund2 &284.3 &  $-0.3$& $\sim 2.8$    & $2.0\pm0.3$  &$ > 7$\tablenotemark{d}     &\citet{ascenso07a}\\     
\enddata		 
\tablecomments{For each cluster, names and Galactic coordinates
 are followed by distances, ages, masses, and references.}
\tablenotetext{a}{ Distance to the Galactic center as given by \citet{eisenhauer05}.}
\tablenotetext{b}{This mass estimate is for the young population.}
\tablenotetext{c}{A mass of 10000 \Msun\ is estimated using a number of 50 stars more massive than 20\Msun\ 
\citep{negueruela08}, and a Salpiter IMF down to 0.8\Msun.}
\tablenotetext{d}{
The cluster mass is likely a lower limit because it was estimated assuming a
distance of 2.8 kpc; recently \citet{naze08} and \citet{rauw07}
reported  a distance of $8.0\pm1.4$ kpc.}
\end{deluxetable}

Recent searches in 2MASS and GLIMPSE surveys have yielded more than 1600
candidate stellar clusters in the Galactic plane.  Even assuming that  50\% of
these candidates may be spurious  \citep{froebrich07}, this number increases of
1/3 the number  of previously known stellar clusters. The properties and spatial
distribution of most of these candidates are not yet known.   Photometric and
spectroscopic follow-up studies are needed in order to confirm the existence of
these candidates, and to analyze their stellar content.

Since infrared cluster searches are less hampered by interstellar extinction,
clusters discovered in  infrared searches are expected to be on average more
distant than the optical ones, enabling us to explore a different region of the
Galaxy (see Fig.\ \ref{xy.ps} and Table \ref{table.massive}). Infrared candidate
clusters may disclose some of the most interesting galactic stellar clusters
based on their location and/or massive stellar content. Studies of the
properties and spatial  distributions of candidate clusters in the  direction of
the inner Galaxy  are of particular interest to verify 1) the paucity  of young
clusters in the central 3 kpc, 2) confirm the possible existence of a stellar
ring  surrounding the central Bar \citep[e.g.][]{ng94}, 3)  locate other
particular structures.  Among the candidate clusters, some of the most massive
ones could be hidden behind walls of interstellar extinction, as demonstrated by
recent infrared discoveries of young massive clusters
\citep{davies07,figer06,messineo08,borissova08}.

Inspired by the above mentioned motivations we have started a long term
project to study candidate clusters. Here, we reported on new
spectroscopic observations of the  brightest members of three inner Galactic
candidate clusters: GLIMPSE20, GLIMPSE13, and Quartet. 

The Quartet cluster is a young cluster (4 to 8 Myr) with a total mass of about
1300--5200 \Msun. Several massive rare stars are detected among the cluster
members: a WC9 star, two Ofpe/WN9 stars, and  two  OB stars.  The Quartet
cluster has an interstellar extinction of \Ak$=1.6\pm0.4$ mag, and is located in
the direction of the HII region G024.83+00.10, which is at a kinematic distance
of 6.1 kpc.  The Quartet cluster represents another episode of star formation in
a region of particular interest, at l=25\degr\ and at a galactocentric distance
of 3.5 kpc, i.e. at the near endpoint of the Galactic Bar
\citep[e.g.][]{habing06}.  This is the same location as that of the two massive
RSG1 and RSG2 clusters \citep{figer06,davies07}.

GLIMPSE20 is a young cluster (6--8 Myr), with a mass of $\sim3400$ \Msun.
It contains a YSG and an early WC star. GLIMPSE20  is located at a
heliocentric distance of about 3.5 kpc, at an \Aks$=1.3\pm0.4$ mag,
and  is probably related to the Sagittarius arm.

The GLIMPSE13 cluster shows a clump of infrared bright stars.   Their infrared
spectra reveal CO band heads at 2.29 \um\ in absorption. The equivalent widths
of the CO band suggest that these stars are giant stars.  Simulations of simple
stellar populations suggest GLIMPSE13 is an older cluster (50--100 Myr) with a
mass of the order of $6700$ \Msun, at a distance of about 3 kpc and \Aks=1.3
mag.  Only a few other known clusters exist with similar ages and masses
\citep{piskunov08,frinchaboy08}.

GLIMPSE20 and Quartet contain several massive stars  in rare short-lived
phases, e.g. WC and Ofpe/WN9 stars.  There are only two other stellar
clusters known to contain   Ofpe/WN9 stars (Quintuplet and the Galactic
center cluster). About a dozen clusters are known to contain WR stars
\citep{kurtev07,massey01}.  The presence of an associated stellar
cluster enables us to measure the initial mass of these evolved stars.
GLIMPSE13 contains stars with masses of $7\pm2$ \Msun, in a transitional
region between massive stars that will end their life exploding as a
supernovae and stars that will cool down as a white dwarf. Their stellar
content deserves further studies.

All three clusters are massive. Massive clusters of different ages  can provide
information on the  timescale of dynamical cluster evolution (cluster survival,
evaporation rate) \citep{gieles08,lamers05,piskunov08}.
The three clusters are located at distances  larger than  those  of  clusters
detected in the optical (see Fig.\ \ref{xy.ps}).  Studies of infrared candidate
clusters appear to be a  promising way of mapping the inner Galaxy. New candidate
clusters will soon be available from the UKIDSS survey of the Galactic plane, as
well as from the  future VISTA survey.

\acknowledgments {The material in this work is supported by NASA under
award NNG 05$-$GC37G, through the Long-Term Space Astrophysics
program. This research was performed in the Rochester Imaging Detector
Laboratory with support from a NYSTAR Faculty Development Program
grant.  This publication makes use of data products from the Two
Micron All Sky Survey, which is a joint project of the University of
Massachusetts and the Infrared Processing and Analysis
Center/California Institute of Technology, funded by the National
Aeronautics and Space Administration and the NSF. This research has
made use of Spitzer's GLIMPSE survey data, the Simbad and Vizier
database. Based on observations collected at the European Southern
Observatory, La Silla, Chile, within the programme 079.D$-$0807(A).
We thank the referee  for his insightful comments.}

\end{document}